\documentclass[11pt]{article}
\usepackage{moriond,epsfig}

\def\NPB{{\em Nucl. Phys.} B}

\def\be{\begin{equation}}
\def\ee{\end{equation}}
\def\bea{\begin{eqnarray}}
\def\eea{\end{eqnarray}}

\newcommand{\dstlnu}{\mbox{$D^*\ell\nu$}}

\newcommand{\dstxlnu}{\mbox{$D^* X\ell\nu$}}

\newcommand{\vcb}{\mbox{$|V_{cb}|$}}
\newcommand{\vub}{\mbox{$|V_{ub}|$}}

\newcommand{\barb}{\bar{B}}

\newcommand{\mb}{m_B}

\newcommand{\gev}{$\:{\rm GeV}$}

\newcommand{\etal}{{\it et al.}}
\newcommand{\PRL}[3]{{Phys. Rev. Lett.} {\bf #1}, #2 (#3)}
\newcommand{\PRD}[3]{{Phys. Rev. D} {\bf #1}, #2 (#3)}
\newcommand{\PLB}[3]{{Phys. Lett. B} {\bf #1}, #2 (#3)}

\newcommand{\ARN}[3]{{Annu. Rev. Nucl. Part. Sci.} #1, #2:#3}

\begin{document}
\vspace*{4cm}
\title{RECENT RESULTS FROM CLEO}

\author{H. SCHWARTHOFF}

\address{Cornell University, Wilson Laboratory, Dryden Road,\\
Ithaca, NY 14853, USA}

\maketitle\abstracts{
Between 1990 and 2001, the CLEO II/II.V/III detectors at the Cornell 
Electron Storage Ring (CESR) have recorded over of 34 million 
$B$ decays and nearly 60 million charm decays. A selection of the latest
results in the electroweak sector from these data sets has been presented.
The focus is placed on the determination of the CKM matrix elements \vub, and \vcb,
and how the measurement of the $b\to s \gamma$ photon spectrum can be
used to reduce theoretical uncertainties.}

\section{Introduction}
\subsection{The CLEO experiment}\label{subsec:cleo}

The data presented here have been recorded with the CLEO detector at the
symmetric 10.58\,GeV $e^+e^-$ collider CESR at Cornell University, NY, USA.
The center of mass energy is chosen to produce $\Upsilon$(4S) mesons in resonance,
which decay predominantly into a pair of $B$ mesons. The rest of the collisions produce
non-resonant $q\bar{q}$ pairs.
One third of the data were recorded at a center of mass energy of
approximately 60\,MeV below the resonance ({\em off-resonance}), to be used
for background studies.

During the CLEO II and II.V phases of the experimental program almost 10 Million
$B\barb\ $ pairs have been recorded in the years 1990--2000. This corresponds to
an on-resonance luminosity of $\approx 9\,fb^{-1}$ and is the subject of this proceedings.
Between 2000 and 2001, an additional $7\,fb^{-1}$ (on-resonance) have been recorded with
an upgraded detector, CLEO III. Analysis on this data is in progress, and the results are
expected to considerably reduce experimental errors from many previous measurements.

\subsection{The CKM matrix}

\begin{figure}
\psfig{figure=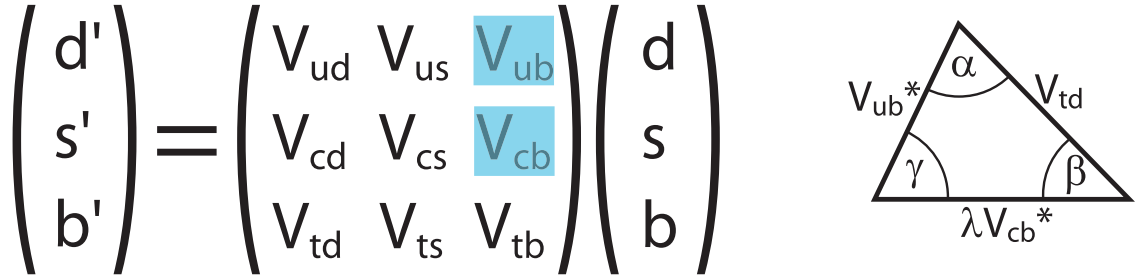,height=2.2in}
\caption{Left: CKM matrix. The elements that are highlighted are subject of this
presentation. Right: Unitarity triangle. Only the parameters relevant for
this presentation are indicated. $\lambda$ is the experimentally well measured sine
of the Cabibbo angle.
\label{fig:ckmut}}
\end{figure}

The Cabibbo-Kobayashi-Maskawa matrix\,\cite{ckm} describes the weak couplings between
the quarks in the Standard Model through 9 fundamental parameters, see Fig.\ 
\ref{fig:ckmut} left. Measurements of those parameters allow one to test existing theories
and can give constraints on physics phenomena beyond the Standard Model.
This Model requires that the matrix be unitary, which is commonly displayed
through the ``Unitarity Triangle'',\cite{unitaritytriangle} see Fig.\ 
\ref{fig:ckmut} right. The knowledge of all three angles of the triangle
is also likely to establish a clear understanding of CP violation in the $b$ quark sector.

The CKM matrix element \vcb\ sets the length of the base of
this triangle. It has been previously measured with an accuracy of
$\approx$\ 5\,\%.\cite{pdg} The element \vub\ bears the largest uncertainty
for measurement of a second side, $\approx$ 30\,\%.\cite{pdg} We will show
that these errors have been  considerably reduced using new analysis techniques
and larger data samples.

\section{How to measure \vcb\ and \vub}

The measurements of \vcb\ and \vub\ are carried out using the semileptonic quark decays
$b\to c\ell\nu$ and $b\to u\ell\nu$, where $\ell$ is any lepton -- typically $e$ or $\mu$.
These decays provide a clear experimental
signature (a single high energy lepton), and are well accessible to theoretical
calculations. Heavy Quark Effective Theory\,\cite{hqet} (HQET) has proven to be a
phenomenological approach that is able to provide quantitative predictions for
observables with uncertainties that lie within the experimental errors.
The method initially calculates an observable assuming the heavy quark in the decay
to be infinitely heavy. It then determines corrections in powers of $1/\mb$, where
$\mb$ is the $B$ meson mass. Two approaches are pursued:

\subsection{Exclusive measurements}

A suitable decay channel is chosen for exclusive measurements, relying on the ability
to fully reconstruct one of the $B$ mesons into known final states. To derive the
CKM matrix element, knowledge of the hadronic form factor that is specific for the
decay is needed. The form factors have not been measured with sufficient accuracy,
and they can usually not be calculated over the full kinematic range. For \vcb,
HQET can access them in the range of maximum momentum transfer between the heavy
quark and the $\ell\nu$ system. Using the relation d$\Gamma_{\mbox{exclusive}}/dw
= \mbox{const} \times F(w)|V_{cb}|$, where $F(w)$ is the form factor as function
of the kinematic variable $w$, we can determine $|V_{cb}|$ by extrapolating the
measured decay width $\Gamma_{\mbox{exclusive}}$ to $w = 1$. This corresponds to the
case of maximum momentum transfer. HQET works well in exclusive $b\to c$ decays,
because $1/M_B$ is small and $1/M_{D^{(*)}}$ is also small. For \vub, one needs another
way to compute the form factor, e.g. Lattice QCD.

\subsection{Inclusive measurements}

An inclusive analysis determines the branching fraction ${\cal B}(B\to X_q\ell\nu$),
where $X_q$ is any hadron produced in the quark decay $b\to q\ell\nu$ ($q = c,u$). The
experimental identification of this hadron is subject of detailed investigation and
a source of considerable systematic uncertainties. Using the measured branching fraction,
one can determine the CKM matrix element $|V_{qb}|$ from expressions provided by HQET.
This requires to make the assumption of {\em quark-hadron duality}, which means that
it is justified to use calculations made on the quark level to describe hadronic
processes, if we integrate over a sufficiently large number of hadronic final states
and sufficiently large fraction of phase space. In considering the analysis results,
it is as well possible to judge the extent to which this is true. As we will point
out below, we have used experimentally measured parameters from another CLEO analysis,
the measurement of the $b\to s\gamma$ photon spectrum, to minimize the theoretical
uncertainties.

\section{CLEO measurement results}

\subsection{$B\to$\dstlnu}

The branching fraction of $B\to$\dstlnu\ is the highest among the exclusive branching
fractions of the $B$ meson, $\approx 5\%.$\cite{pdg} In the corresponding recent CLEO
analysis,\cite{dstarlnu} 3.3 Million $B\barb\ $ from the CLEO II experiment have been
used. The event reconstruction identifies the desired signatures
by searching for the decay of charged and neutral $D^*$ mesons to a $D\pi$ pair.
The background spectra are obtained from off-resonance data and, to a small part, 
Monte Carlo simulation. The relative contributions of $B\to\dstlnu$ and $B\to\dstxlnu$
are fitted to the data.

The differential decay width d$\Gamma_{\mbox{exclusive}}/dw$ for the decay $B\to\dstlnu\ $
can be expressed as a function of $\vcb F(w)$, where w is a kinematic variable that is
equal to the scalar product of the 4-vectors of the $B$ meson and the $D^*$. It ranges from
$w = 1$ (max.\ momentum transfer, {\em zero recoil}) to $w \approx 1.5$. The
extrapolation of $\vcb F(w)$ to $w = 1$ yields $\vcb F(1) = 0.0431 \times
(1\pm 3\% \pm 4.2\%)$, where the first error is statistical, and the second
systematic. Using $F(1) = 0.919^{+0.03}_{-0.035}$ from a recent
lattice QCD calculation,\cite{latticef1} we determine
$\vcb = 0.047 \times (1 \pm 3\% \pm 4.2\% \pm 3.8\%)$, where the third error comes
from $F(1)$. This result is the single most precise \vcb\ exclusive measurement to date.

\subsection{$B\to\rho,\pi\,\ell\nu$}

The exclusive analysis to determine \vub\ investigates the decay channels
$B\to\rho,\pi,\omega \,\ell \nu$. By the time of the Moriond 2002 conference it has
been ongoing both utilizing the full CLEO II and CLEO III datasets, relying on a
refined procedure for neutrino reconstruction.\cite{neutrino}
The background is determined in
a similar way as for the $B\to$\dstlnu\ work. Since the form factors for the
considered channels are not as well known, the analysis will actually be able to provide
constraints on the existing form factor models. A \vub\ measurement with a total
uncertainty of $\approx 15\%$ is expected to be available by summer 2002. It will
supercede previous CLEO measurements,\cite{vub1,vub2} which are accurate to
$\approx 20\%$.

\subsection{$B\to s\gamma$}

\begin{figure}
\hspace{3cm}
\psfig{figure=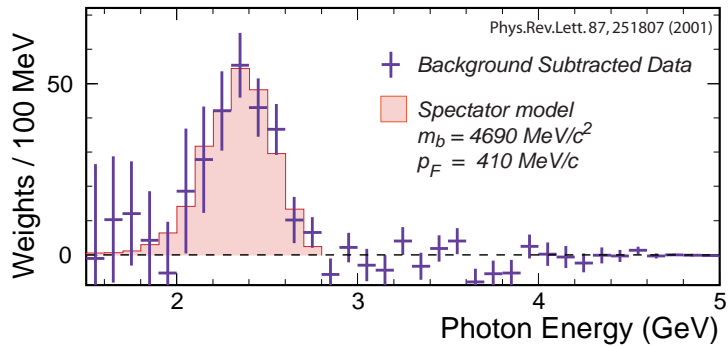,height=5cm}
\caption{Photon spectrum for $b\to s \gamma$. The background subtracted data
(data points) have been fitted to an HQET expression (Spectator model) with the
mean energy as free parameter.
\label{fig:btosgammaspectrum}}
\end{figure}

The recent CLEO analysis\,\cite{btosgamma} measuring the properties of the inclusive decay
$B\to X_s\gamma$ has been able to supply a valuable input for all other inclusive
semileptonic channels: the $b\to s\gamma$ photon spectrum. The previous CLEO
measurement\cite{btosgammaold} of the inclusive $b\to s\gamma$ branching fraction was of
significant interest to the theoretical physicist, but the resulting photon spectrum
was not sufficiently precise for comparison with HQET predictions. After using the
full CLEO II and II.V dataset and improving the $B\barb\ $ background suppression
considerably, the resulting spectrum (see fig.\ \ref{fig:btosgammaspectrum}) allows
the determination of the HQET parameter $\bar{\Lambda}$.

This parameter is a measure for the light degrees of freedom in the $B$ meson.
The mean photon energy can be expressed as follows:
$<E_\gamma> = 1/2M_Bf(\bar{\Lambda}/M_B)$, where $M_B$ is the $B$ meson mass, and
$f(\bar{\Lambda}/M_B)$ is a known function that is (approximately) independent of
the $B$ decay channel. We find $\bar{\Lambda} = 0.35 \pm 0.08 \pm 0.1$\gev.

\subsection{$B\to X_c\ell\nu$}

The mean of of the charmed hadron mass spectrum in the inclusive channel
$B\to X_c\ell\nu$ has been predicted by HQET:\cite{falk}
$<M_X^2 - M_D^2> = f(\bar{\Lambda}, \lambda_1)$, where $f$ is a given function of the
above determined parameter $\bar{\Lambda}$, and a second HQET paremter, $\lambda_1$,
which is a measure for the average momentum of the $b$ quark in the $B$ meson.
The expression gives the mean hadronic mass with respect to the $D$ meson mass $M_D$
and has been obtained to order $1/M_B^3$, in the $\overline{MS}$ renormalization scheme.
It is thus possible to derive $\lambda_1$ from the hadron spectrum measurement.
In the corresponding CLEO analysis,\cite{btoxclnu} a powerful background suppression
is employed by only using lepton momenta above 1.5\gev\ and applying
the CLEO neutrino reconstruction method.

\begin{figure}
\hspace{4cm}
\psfig{figure=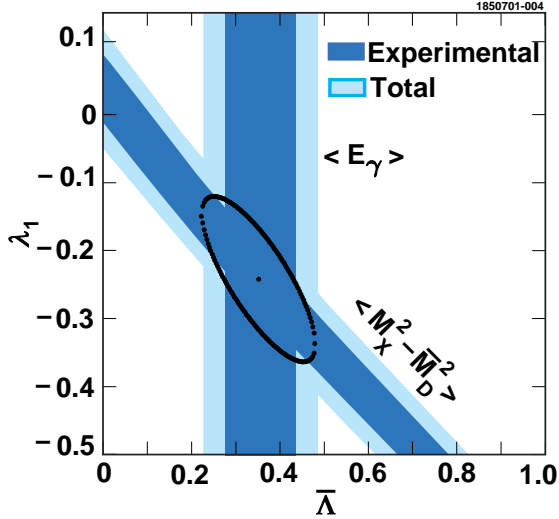,height=7cm}
\caption{Determination of the HQET parameters $\bar{\Lambda}$ and $\lambda_1$ by
combining the measurement results for the $b\to s\gamma$ photon spectrum
($<E_\gamma>$) and the $B\to X_c\ell\nu$ hadronic mass spectrum ($<M_X^2 - M_D^2>$).
\label{fig:hadronicmoments}}
\end{figure}

Combining this measurement with the $\bar{\Lambda}$ (see fig.\ \ref{fig:hadronicmoments})
result from the $b\to s\gamma$ analysis allows the determination of
$\bar{\Lambda} = 0.35 \pm 0.07 \pm 0.1$\gev, and
$\lambda_1 = -0.236 \pm 0.071 \pm 0.078$\gev.

The determination of \vcb\ is now possible using an expression\cite{gammasl}
that links the full semileptonic decay width $\Gamma_{sl}$ to \vcb\ times a function
$h(\bar{\Lambda},\lambda_1)$.
$\Gamma_{sl} = (0.427\pm 0.02) \times 10^{-10}$
can be obtained from previous CLEO measurements of
${\cal B}(B\to X_c\ell \nu)$,\cite{bbrfrac}
the $B$ meson lifetime $\tau_{B^\pm},$\cite{pdg}
and the ratio of charged to neutral $B$ meson production rates
$f_{+-}/f_{00}$.\cite{sylvia}
The final result is $\vcb = 0.0404 \times (1 \pm 2.3\% \pm 1.3\% \pm 2\%)$. The total
error of $3.2\,\%$ is the lowest for any measurement. It needs to be mentioned again
that the assumption of quark-hadron duality is made for this conclusion.

\subsection{$B\to X_u\ell\nu$}

Measuring \vub\ has been historically difficult, because of theoretical uncertainties.
One experimental reason is the fact that semileptonic $B$ decays are highly dominated 
by $b\to c$ transitions, which are favored over $b\to u$ by two orders of magnitude.
A technique that has been applied in earlier works\,\cite{oldbtou} exploits
the kinematic situation of the decay by measuring the lepton momentum spectrum
in the kinematic end region towards higher lepton momenta, where the $b\to c$
decays are suppressed due to momentum conservation. For the current
analysis\,\cite{btoulnu} this method has been refined through the improvement of the
$B\to X_c\ell\nu$ Monte Carlo modeling, which could be achieved using recent form
factor measurements and a more precise HQET description. Also, the continuum
suppression was strongly improved through the use of a neural net, which lead to
a reduction in model dependence.

The full decay width for $B\to X_u\ell\nu$ is required for the \vub\ determination.
Since we only measure the spectrum in the end point region, we determine the measured
fraction of the distribution f$_u$ by fitting the measured $X_u\ell\nu$ lepton energy
spectrum to a nonperturbative shape function.\cite{btosgammashape}
This shape function can be obtained from the photon energy spectrum in the $b\to s \gamma$
analysis.\cite{btosgamma} We find the full decay width by dividing the measured
partial width by f$_u$, then extract \vub\ using the following equation:\cite{vubeq}
\be
\vub = (3.07\pm 0.12)\times 10^{-3}\times (\frac{{\cal B}(B\to \mbox{X}_ue\nu)
\times 1.6\,\mbox{ps}}{0.001 \tau_B})^\frac{1}{2}.
\label{eq:vub}
\ee
The choice of the optimal momentum interval for \vub\ is a compromise between the
limited knowledge of the $B\to X_c\ell\nu$ background and the uncertainty in f$_u$.
The results for various intervals are consistent with each other. We choose the
lepton momentum interval between 2.2 and 2.6\gev: $\vub = 0.00408 \times (1 \pm 8.3\%
\pm 10.8\% \pm 3.9\% \pm 5.9\%)$. Assuming that quark-hadron duality holds for this
analysis, we consider this result to be the best \vub\ determination to date.

\section*{Acknowledgments}
The uniqueness of the Moriond conference series has provided a wonderful experience
in international research exchange. I also acknowledge the National Science Foundation
for supporting conference participation.

\end{document}